\documentstyle[aps,preprint,graphics]{revtex}

\newcommand{\f}[1]{$\mathrm{F_{#1}}$}
\tightenlines
\begin{document}
\draft
\title{Quantitative investigation of the mean-field scenario for the
structural glass transition from a schematic mode-coupling analysis of
experimental data}
\author{V.~Krakoviack\thanks{E-mail: krako@lcp.u-psud.fr} and
C.~Alba-Simionesco\thanks{E-mail: chalba@lcp.u-psud.fr}}
\address{LCP,UMR 8611, b\^atiment 490, Universit\'e Paris Sud, F-91405 Orsay}
\maketitle

\begin{abstract}
A quantitative application to real supercooled liquids of the mean-field
scenario for the glass transition ($T_g$) is proposed. This scenario,
based on an analogy with spin-glass models, suggests a unified picture
of the mode-coupling dynamical singularity ($T_c$) and of the entropy
crisis at the Kauzmann temperature ($T_K$), with $T_c>T_g>T_K$. Fitting
a simple set of mode-coupling equations to experimental light-scattering
spectra of two fragile liquids and deriving the equivalent spin-glass
model, we can estimate not only $T_c$, but also the static transition
temperature $T_s$ corresponding supposedly to $T_K$. For the models and
systems considered here, $T_s$ is always found above $T_g$, in the fluid
phase. A comparison with recent theoretical calculations shows that this
overestimation of the ability of a liquid to form a glass seems to be a
generic feature of the mean-field approach.
\end{abstract}


Despite considerable experimental and theoretical efforts, understanding
the dynamics of supercooled liquids and the related phenomenon of glass
transition remains a challenging problem of classical statistical
mechanics \cite{ang95sci}.

Recently, Kirkpatrick \emph{et al.} \cite{kirketal,kirketal2}
conjectured that generalized spin-glass models with discontinuous
one-step replica symmetry breaking (1RSB) transitions \cite{sgtab} (like
the spin-glass with $p$-spin interactions ($p>2$) or the $q$-state Potts
glass ($q>4$)) could be relevant models for the description of the
structural glass transition. Essentially two points substantiate this
analogy. Firstly, when considered in the mean-field limit where the
interactions between spins have infinite range, the generalized
spin-glasses display a (static) 1RSB transition to a spin-glass phase at
a temperature $T_s$ that is accompanied by the vanishing of the
configurational entropy similar to the entropy crisis at $T_K$
hypothesized by Kauzmann for glassforming liquids \cite{kauzmann}.
Secondly, in the same mean-field limit the study of the Langevin
dynamics of such models shows that a dynamical transition takes place at
a temperature $T_d$ greater than $T_s$. Above $T_d$, the time evolution
of the spin correlation function is given by a non-linear equation of
the same type as those occurring in the ideal mode-coupling theory (MCT)
for the glass transition of simple liquids, where they describe the time
evolution of density fluctuations \cite{mctrev}, and $T_d$ coincides
with what is called the critical temperature $T_c$ in the context of
MCT. Below $T_d$, the systems display a nontrivial free-energy landscape
that in finite-range models could lead to slow activated dynamics, as
described for instance by the Adam-Gibbs theory \cite{gibbsetal}.

Mean-field models and their extensions to finite-range systems could
thus provide a consistent framework for the study of the liquid-glass
transition, a framework in which the mode-coupling approach finds a
natural place and that catches important aspects of the glass
phenomenology. This approach has recently been very fruitful:
first-principle studies based on the replica method have been proposed
for simple liquid models \cite{mezpar1carfrapar,mezpar2}, and computer
simulation studies of the out-of-equilibrium dynamics of simple liquids
\cite{parisikobbar} show aging behaviors qualitatively similar to the
one displayed by generalized spin glasses \cite{cugkur93prl}.

A question remains, however, elusive. Indeed, it is known that the
dynamical transition predicted by the ideal MCT is not experimentally
observed: it is 'avoided' because of ergodicity restoring processes not
accounted by the theory. In general, it is thought to be replaced by a
smooth crossover regime where the dynamics changes qualitatively
\cite{mctrev}. Because of the similarities between the ideal MCT and the
dynamical aspects of the mean-field theory, the same breakdown is
expected in the latter when more realistic systems with finite-range
interactions are considered. Its influence on the other aspects of the
mean-field picture described above is unknown, but the static transition
is usually assumed to survive and the dynamical freezing is expected to
occur only at the static transition \cite{kirketal2}. This issue clearly
deserves further investigation.

In this Letter, as a step in this direction, we propose a quantitative
study of the above mean-field scenario for \emph{real} supercooled
liquids. By quantitative, we mean that we will extract from experimental
data values for the two characteristic temperatures introduced within
the mean-field approach and discuss these values in relation with known
properties of the investigated systems. At present, no definite method
exists for such a work. We propose here to start from a phenomenological
mode-coupling approach, the so-called schematic approach to experimental
data \cite{albkra95jcp}, which is found to be particularly well suited
for investigating the avoidance of the dynamical transition. Then,
taking advantage of the coincidence between the ideal mode-coupling
equations and the dynamics of some mean-field generalized spin-glass
models, the schematic calculation is rephrased in terms of an effective
spin-glass hamiltonian, whose study allows us to determine, for the
glassforming liquids under investigation, the location of the two
transitions predicted by the mean-field theory.

Schematic models are simple sets of mode-coupling equations
\cite{schem}, which have proven to be useful in testing the MCT on
realistic systems. Indeed, these models can be included within a fitting
procedure of experimental data \emph{over a wide time or frequency
range}, and they allow the calculation of effective mode-coupling
parameters ('vertices') describing the dynamical evolution of a liquid
with varying external conditions
\cite{albkra95jcp,kraalbkra97jcp,glycschem,otpschem}. A major interest
of these models is that they catch the universal features of the
mode-coupling equations, in particular the asymptotic scaling results
valid near the dynamical transition, but are not reduced to them. They
make possible to overcome the difficulties and uncertainties arising
from the need for corrections to the asymptotic critical predictions of
the theory and to take into account the $\alpha$ relaxation as well as
the regime of the high-frequency microscopic excitations. The price to
pay is that these models are somewhat \emph{ad hoc} and might possibly
display non-generic features.

We make use in this Letter of the results obtained from a previous study
\cite{kraalbkra97jcp} of the depolarized light scattering spectra of two
so-called 'fragile' glassformers, CKN and salol \cite{spectra}. Only the
facts relevant to our present calculation are reviewed here and the
reader is referred to the corresponding paper for details. The basis of
our study is the well studied \f{12} model \cite{mctrev}, defined by the
following mode-coupling equation for a correlator $\phi_0(t)$:
\begin{displaymath}
\ddot\phi_0(t)+\nu_0\,\dot\phi_0(t)+\Omega_0^2\,\phi_0(t)+
\Omega_0^2\!\displaystyle{\int^t_0\!\! m_0(t-\tau)\,\dot\phi_0(\tau)\,
\mathrm{d}\tau=0},
\end{displaymath}
with $m_0(t)=v_1\,\phi_0(t)+v_2\,\phi_0^2(t)$, $v_1,v_2\ge 0$. To fit
the experimental light-scattering spectra over the full available
frequency range, in particular to reproduce the shape of the peak in the
THz domain and the extra-intensity of the $\alpha$-peak, we must add a
second correlator $\phi_1$, whose time evolution is given by a similar
equation with $m_1(t)$ simply chosen proportional to $m_0(t)$. The first
correlator $\phi_0(t)$ accounts in an effective way for the slow modes
responsible for the slowing down of the relaxation, possibly leading to
a dynamical transition. The second correlator $\phi_1(t)$ describes the
contribution of additional degrees of freedom that are important in the
light-scattering processes, but whose slow dynamics is dominated by that
of $\phi_0(t)$. The critical slowing down is thus totally driven by the
time evolution of $\phi_0$, whose variation with temperature is entirely
encoded in the $T$-dependence of the 'vertices' $v_1$ and $v_2$ (the
other parameters are taken $T$-independent). Because $\phi_1$ plays no
role in the parametrization of the dynamical evolution, it can be
discarded from the building of the effective generalized spin-glass
model. The latter is obtained remarking that the long-time dynamics
described by the \f{12} equation is \emph{identical} to that of a
mean-field spherical spin-glass with spins interacting via both 2-body
and 3-body random terms, namely
\begin{displaymath}
-\beta\,{H[{\bf s}]}=\sum_{1\leq i_1<i_2\leq N}\ J_{i_1 i_2}\ s_{i_1}\,s_{i_2}\
+\sum_{1\leq i_1<i_2<i_3\leq N} J_{i_1 i_2 i_3}\
s_{i_1}\,s_{i_2}\,s_{i_3},
\end{displaymath}
where the $s_i$'s are spherical spin variables and the couplings $J_{i_1
i_2}$ and $J_{i_1 i_2 i_3}$ are independent gaussian variables with zero
means and variances $\overline{(J_{i_1 i_2})^2}=v_1/N$ and
$\overline{(J_{i_1 i_2 i_3})^2}=2\,v_2/N^2$ \cite{mctpspin}. A
remarkable property of these systems is that the correlation function of
the Hamiltonian,
\begin{displaymath}
\beta^2\,\overline{H[{\bf s}]\,H[{\bf s'}]}=N\left(\frac{v_1}{2}
q_{{\bf s,s'}}^2+\frac{v_2}{3}q_{{\bf s,s'}}^3\right),
\end{displaymath}
where $q_{{\bf s,s'}}={\bf s\cdot s'}/N=\left(\sum_i s_i\,s'_i\right)/N$
is the overlap between the spin configurations $\mathbf{s}$ and
$\mathbf{s'}$, determines completely both their statics and dynamics
\cite{frapar95jp1fr}. With parameters $v_1$ and $v_2$ extracted from a
fit of the dynamics, we are thus able to construct an effective
generalized spin-glass model allowing us to investigate the mean-field
scenario for the glassforming salol and CKN.

\begin{figure}
\vspace{5cm}
\centering
\rotatebox{270}{\scalebox{0.55}{\includegraphics*{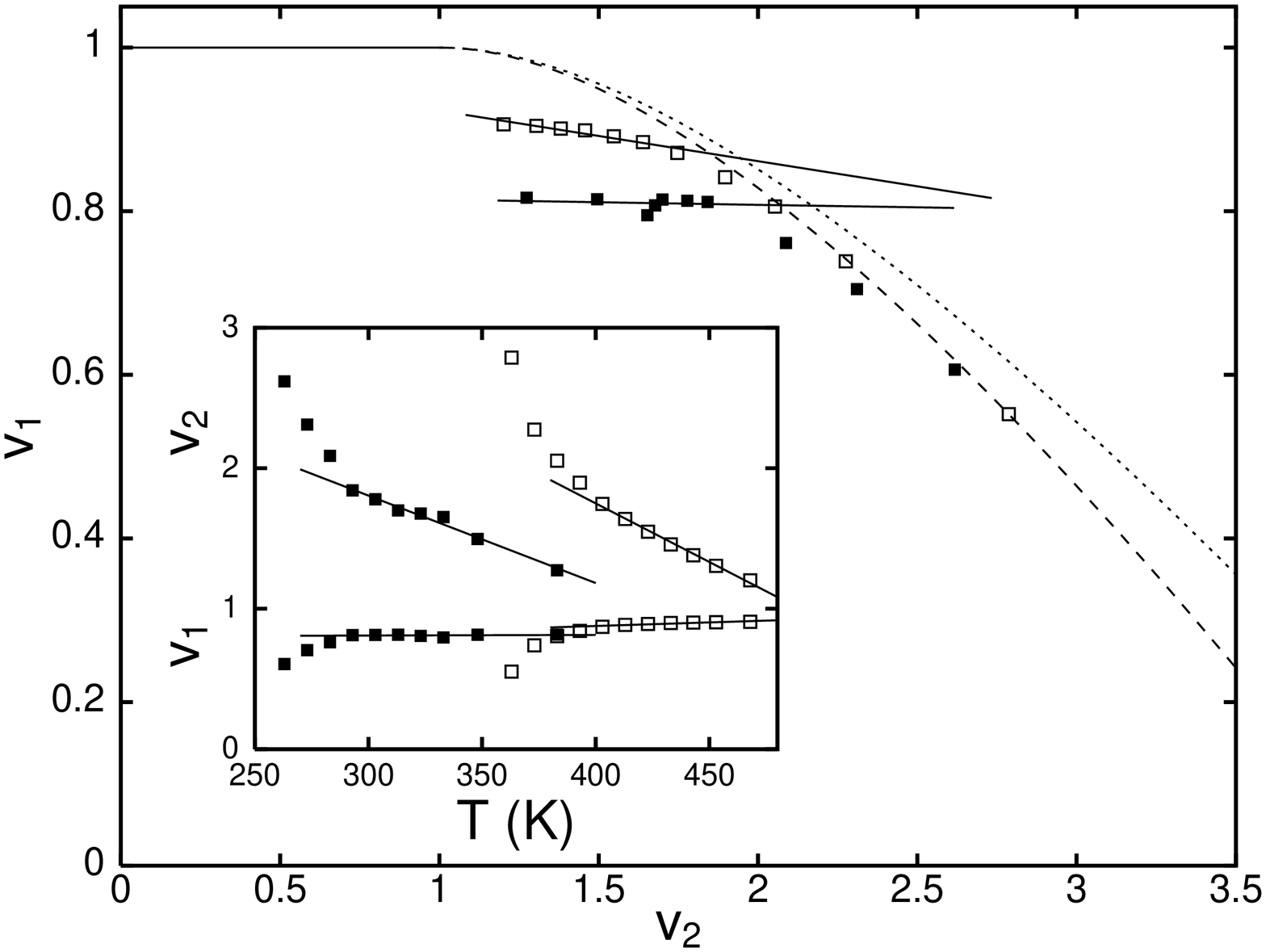}}}
\bigskip
\bigskip

\caption{Phase diagram of the generalized spin-glass model
(continuous line: continuous dynamical and 1RSB transitions, dashed
line: discontinuous dynamical transition, dotted line: discontinuous
1RSB transition) and effective vertices obtained from the MCT fit (open
symbols: CKN, filled symbols: salol). Inset: evolution of the vertices
with temperature. The superimposed straight lines are the linear fits
used for the low-temperature extrapolations.}
\label{vertex}
\end{figure}

The phase diagram of the generalized spin-glass model described above is
easily computed and is plotted on figure \ref{vertex}. The
dynamical-transition line separating the ergodic domain in the vicinity
of the origin from the non-ergodic one in the vicinity of infinity has
two branches: $\{v_1=1,\ v_2\le 1\}$ corresponding to a continuous or
type A transition and $\{v_1=2\,(v_2)^{1/2}-v_2,\ 1< v_2\le 4\}$
corresponding to a discontinuous or type B one, respectively. At the
static level, by using the replica trick to average over the quenched
disorder\cite{sgtab}, the model is exactly solved with a 1RSB ansatz,
which leads to the following free energy density
\begin{displaymath}
\beta f(q,x)=-\frac{v_1}{4}(1-(1-x)q^2)-\frac{v_2}{6}
(1-(1-x)q^3)-\frac{1}{2x}\ln(1-(1-x)q)-\frac{x-1}{2x}
\ln(1-q),
\end{displaymath}
where $q$, the mutual overlap between replicas lying in the same
cluster, and $x$, the cluster size, are variational parameters ($0\leq
q,x \leq 1$), with respect to which the free energy has to be maximized.
Here again, the transition line between the replica-symmetric phase at
small couplings and the 1RSB phase at large couplings has two branches:
a continuous 1RSB transition line coincides with the continuous
dynamical transition line, whereas a discontinuous 1RSB transition line
is found beyond the dynamical discontinuous transition line. It
corresponds to the appearance of a non-zero solution for $q$ with $x=1$
and its equation as a parametric function of $q$ is given by
\begin{displaymath}
v_1=2\frac{2q^2-3q-3(1-q)\ln(1-q)}{q^2(1-q)},\qquad
v_2=3\frac{2q-q^2+2(1-q)\ln(1-q)}{q^3(1-q)}.
\end{displaymath}

On figure \ref{vertex} are also reported the effective vertices obtained
from the experimental data for the two supercooled liquids. They display
two different regimes with varying temperature. At higher temperatures,
they vary linearly and show the apparent evolution of the liquids toward
the dynamical transition expected from MCT. But, above the corresponding
transition temperature, the vertices behavior changes: the transition is
not reached, and the vertices follow the dynamical-transition line
without crossing it. The first regime can be associated with the domain
of validity of the ideal MCT, whereas the second one indicates the
failure of the theory because of the putative onset of activated
processes. Indeed, the MCT states that the dynamics is governed by
vertices which are purely static quantities and thus change smoothly
with external parameters. Even if in the case of schematic models the
connection between the effective vertices and static quantities is
somewhat obscured, it is reasonable to expect smooth variations of the
fitted parameters with temperature. Accordingly, we concentrate in the
following on the high temperature regime and its extrapolation to lower
temperatures. Note that one can avoid the need for extrapolation using a
low frequency cut-off for the lowest temperatures \cite{otpschem}.

With this method, we find the dynamical temperatures $T_d=T_c=257 \pm 5$
K for salol and $T_d=T_c=388 \pm 5$ K for CKN, in good agreement with
previously determined values of the mode-coupling transition
temperature. By construction, the mode-coupling transitions of the
liquids and the dynamical transitions of the effective disordered
systems are identical.

We now turn to the static calculation. We are interested in the
discontinuous 1RSB transition temperature, since it is thought to
describe the entropy crisis associated with the resolution of the
Kauzmann paradox. Indeed, as first pointed by Kauzmann, because the heat
capacity of a supercooled liquid is substantially greater than that of
the underlying crystalline solid, reasonable extrapolations of the
liquid entropy below the glass transition seem to cross the entropy of
the crystal at a non-zero temperature $T_K$. To solve this paradox,
following Gibbs and Di Marzio \cite{GibDi}, it is sometimes postulated
that, below $T_g$, a second-order transition to an ideal glassy state
should exist, at which the configurational contribution to the entropy
of the liquid vanishes. This mechanism is precisely at work in the
mean-field models at a discontinuous 1RSB transition. Between the
dynamical and static transitions, one finds that the Gibbs measure is
dominated by a number of states exponentially large in $N$, leading to a
finite configurational entropy density (defined as the logarithm of that
number of states divided by $N$). At the static transition, this
configurational entropy density vanishes and stays zero in the
low-temperature phase.

\begin{figure}
\centering
\scalebox{.55}{\includegraphics*[3cm,8.5cm][24cm,35cm]{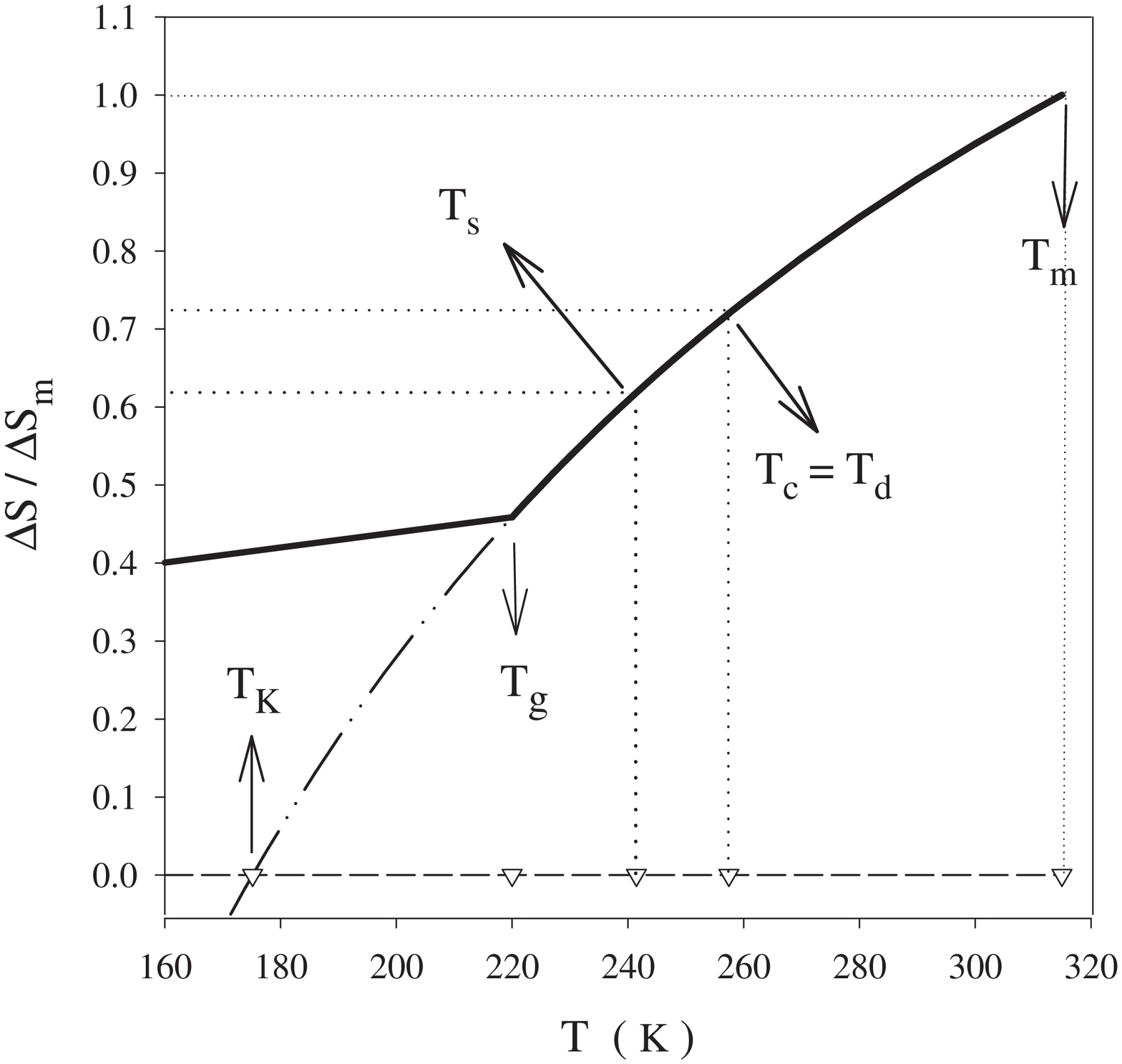}}
\bigskip
\bigskip

\caption{Characteristic temperatures calculated within
the mean-field scenario for salol, $T_c=T_d$ and $T_s$, represented on a
configurational entropy plot (normalized to the entropy of melting
$\Delta S_m$). Entropy data taken from Ref. \protect\cite{hiroki}. Also
shown are the calorimetric glass transition temperature $T_g$, the
melting temperature $T_m$, and the extrapolated Kauzmann temperature
$T_K$. The predicted static transition $T_s$ occurs well above $T_g$.}
\label{entropy}
\end{figure}

From the above analysis, we find $T_s=242 \pm 5$ K for salol and
$T_s=376 \pm 5$ K for CKN. These values, which are very close to $T_d$,
have to be compared with the experimental calorimetric glass transition
temperatures ($T_g=220$ K for salol, $T_g=333$ K for CKN) and the
empirically determined Kauzmann temperatures ($T_K=175$ K for salol
\cite{ricang98jcp}, no value for CKN because the crystalline phase is
unstable): $T_s$ is found in both cases above $T_g$, \emph{i.e.} still
in the liquid phase ! All these characteristic temperatures are plotted
in figure \ref{entropy} in the case of salol: the total configurational
entropy decreases by only $10\%$ between $T_d$ and $T_s$.

We have investigated the effect of minor modifications of the schematic
calculation on our results, namely changing the expression of the
calculated susceptibility as a functional of $\phi_0$ and $\phi_1$ (each
can enter linearly or quadratically in the susceptibility expression)
and/or changing the second memory function to
$m_1(t)=r\,\phi_0(t)\,\phi_1(t)$. The resulting fits are all of equally
good quality, and the corresponding vertex trajectories, although
slightly different, agree with the previous values for both the two
transition temperatures and the vertices at the dynamical transition. As
these changes merely affect $\phi_1$, this consistency validates our
assumption of neglecting it for building the effective model. For more
drastic changes to the model, in which the \f{12} equation is replaced
by another one (for instance, we tried the \f{13} and \f{29} models
where $m_0(t)=v_1\,\phi_0(t)+v_3\,\phi_0^3(t)$ and
$m_0(t)=v_2\,\phi_0^2(t)+v_9\phi_0^9(t)$, respectively), we have not
been able to fit satisfactorily the experimental susceptibilities.
Indeed, the fitted curves failed to reproduce the location of the
susceptibility minimum, its shape, or else the position of the
$\alpha$-peak. As these features are crucial for the characterization of
the dynamics within the MCT framework, the results, in qualitative
agreement with those obtained with the \f{12} model, do not appear
reliable enough for quantitative purpose. The origin of this failure is
in general unclear, but in some cases can be related to non-generic
features of a given schematic model (for instance, the \f{13} model
displays an $\mathrm{A}_3$ singularity very close to the calculated
vertex trajectories). This limitation seems to be severe for our static
calculation, as it is known from the study of mean-field models like
spin-glasses with $p$-spin ($p>2$) interactions \cite{crisom92zpb} or
the Potts glass \cite{desparrit}, that the larger the asymptotic value
of the correlation function at the dynamical transition, the larger the
ratio $T_d/T_s$, and the \f{12} model only allows for small values of
the former. But we stress that, in our model, the temperature enters in
a different way, only through the effective vertices whose dependence
comes out directly from the fitting procedure to dynamic
light-scattering susceptibilities. There is, thus, no built-in closeness
of $T_d$ and $T_s$ in our work. A related remark is the independence of
our results with respect to changes in the way $\phi_0(t)$ enters the
calculated susceptibility, thereby showing that its infinite time limit
is not a sensitive parameter in the study.

Within our phenomenological implementation of the mean-field approach,
we find thus that this theory seems to overestimate notably the tendency
of a supercooled liquid to form a glass. This overestimation shows up at
two levels ($T_d>T_g$ on one hand, $T_s>T_g$ on the other) with clearly
different implications. As stated in introduction, this result is
expected from the dynamical side of the theory, because of its closeness
with the ideal mode-coupling theory, whose inadequacy at low temperature
is well known, and of the need to take into account corrections to
mean-field in finite dimension \cite{kirketal2}. From this point of
view, this conclusion is not new. What is more unexpected is that the
obtained static temperature values are clearly located in the fluid
domain and do not seem to be associated to any change of behavior of the
studied systems. These results are thus inconsistent with the arguments
prescribing that, going beyond mean-field, the real dynamical transition
should occur at $T_s$ for finite dimensional systems \cite{kirketal2}.

Whether the found overestimation has to be ascribed to an inadequacy of
our simple phenomenological approach or more fundamentally to the theory
itself can not be answered here \emph{a priori}. We can nevertheless
remark that theoretical studies of simple liquid models based on the
replica method \cite{mezpar1carfrapar,mezpar2} tend to support our
observations and conclusions. Indeed, one finds in general that the
location of the static transition agrees well with the glass transition
found by computer simulation, \emph{i.e.} obtained with large quenching
rates and on short observation times. This agreement implies thus an
overestimation of the ability of the liquids to freeze into a glass on a
macroscopic time scale, as found in our calculation. Concerning the
closeness of the dynamical and static transitions, our results can only
be compared consistently with theoretical approaches allowing one to
compute the locations of both transitions within the same framework.
This is the case in the papers of Ref. \cite{mezpar1carfrapar} only, in
which the transitions are found very close to each other, just as we
find here.

To summarize, we have investigated a potential application of the
mean-field scenario for the liquid-glass transition to \emph{real}
supercooled liquids by constructing an effective generalized spin-glass
model whose dynamics reproduces the standard mode-coupling equations
used in the ideal MCT of glassforming liquids. We have been able to
determine from experimental data for two fragile glass-formers both the
dynamical and static transitions, which are found to be rather close and
both located in the liquid phase. This result, in qualitative agreement
with recent theoretical studies, shows that the theory apparently
overestimates the ability of a supercooled liquid to freeze into a
glass, at least in its simple implementation considered here. Whether
this deficiency could be cured by employing a more sophisticated (but
yet unknown) version of the mean-field approach to real supercooled
liquids or would require a non mean-field treatment accounting
explicitly for activated dynamics remains an open question.
\enlargethispage{0.2cm}

We are grateful to Prof.\ W.~G\"{o}tze and Dr.\ M.~Fuchs for providing
us the codes for some parts of the calculations. We are also indebted to
Prof.\ H.Z.~Cummins and Dr.\ G.~Li for the use of their experimental
data. Dr.\ G.~Tarjus and Dr.\ J.~Kurchan are particularly acknowledged
for fruitful discussions.

\end{document}